\documentclass[conference]{IEEEtran}
\IEEEoverridecommandlockouts
\usepackage{cite}
\usepackage{multirow}
\usepackage{amsmath,amssymb,amsfonts}
\usepackage{algorithmic}
\usepackage{graphicx}
\usepackage{textcomp}
\usepackage{svg}
\usepackage{acronym}
\usepackage[nopatch=footnote]{microtype}
\acrodef{EDFA}{Erbium-Doped Fiber Amplifier}
\acrodef{VOA}{Variable Optical Attenuator}
\acrodef{SELU}{Scaled Exponential Linear Unit}
\acrodef{NN}{Neural Network}
\acrodef{ReLU}{Rectified Linear Unit}
\acrodef{MSE}{Mean Squared Error}
\acrodef{MAE}{Mean Absolute Error}
\acrodef{TL}{Transfer Learning}
\acrodef{ML}{Machine Learning}
\acrodef{SNN}{Self Normalizing Neural Networks}
\acrodef{WDM}{Wavelength Division Multiplexing}
\acrodef{DWDM}{Dense Wavelength Division Multiplexing}
\acrodef{WSS}{Wavelength Selective Switch}
\acrodef{ITU}{International Telecommunication Union}
\acrodef{ROADM}{Reconfigurable Optical ADD-DROP Multiplexer}
\acrodef{OCM}{Optical Channel Monitor}
\acrodef{PD}{Photo Diode}
\acrodef{PM}{Power Monitor}
\acrodef{OSNR}{Optical Signal-to-Noise Ratio}
\acrodef{GFF}{Gain Flattening Filter}
\acrodef{ILA}{In-Line Amplifier}
\acrodef{PD}{Photo-Diode}
\acrodef{LR}{Learning Rate}
\acrodef{ASE}{Amplified Spontaneous Emission}

\usepackage{xcolor}
\def\BibTeX{{\rm B\kern-.05em{\sc i\kern-.025em b}\kern-.08em
    T\kern-.1667em\lower.7ex\hbox{E}\kern-.125emX}}
\begin{document}

\title{Transfer Learning for EDFA Gain Modeling: A Semi-Supervised  Approach Using Internal Amplifier Features\\

\thanks{Supported by grants from SFI: 12/RC/2276\_p2, 18/RI/5721, and 13/RC/2077\_p2}
}

\author{\IEEEauthorblockN{Agastya Raj}
\IEEEauthorblockA{\textit{Computer Science and Statistics, CONNECT} \\
\textit{Trinity College Dublin}\\
RAJAG@tcd.ie}
\and
\IEEEauthorblockN{Dan Kilper}
\IEEEauthorblockA{\textit{Engineering, CONNECT} \\
\textit{Trinity College Dublin}\\}
\and
\IEEEauthorblockN{Marco Ruffini}
\IEEEauthorblockA{\textit{Computer Science and Statistics, CONNECT} \\
\textit{Trinity College Dublin}\\ }

}

\maketitle

\begin{abstract}

The gain spectrum of an Erbium-Doped Fiber Amplifier~(EDFA) has a complex dependence on channel loading, pump power, and operating mode, making accurate modeling difficult to achieve. Machine Learning~(ML) based modeling methods can achieve high accuracy, but they require comprehensive data collection. We present a novel ML-based Semi-Supervised, Self-Normalizing Neural Network~(SS-NN) framework to model the wavelength dependent gain of EDFAs using minimal data, which achieve a Mean Absolute Error~(MAE) of 0.07/0.08 dB for booster/pre-amplifier gain prediction. We further perform Transfer Learning~(TL) using a single additional measurement per target-gain setting to transfer this model among 22 EDFAs in Open Ireland and COSMOS testbeds, which achieves a MAE of less than 0.19 dB even when operated across different amplifier types. We show that the SS-NN model achieves high accuracy for gain spectrum prediction with minimal data requirement when compared with current benchmark methods.

\end{abstract}


\section{Introduction}

Optical networks play a crucial role in supporting new services, due to their ability to meet the high bandwidth, low latency, and reliability requirements~\cite{5G_review}. In addition, they are increasingly important for supporting access and metro optical convergence due to their ability to unify different network layers, enhance efficiency, and meet the growing demands for high-speed connectivity~\cite{Convergence}. To transfer data over long distances and across access and metro domains, optical networks are amplified with \acp{EDFA} to boost optical signals to overcome fiber and link losses. The end-performance metrics such as \ac{OSNR}, depends on the accumulated noise through the network. Thus, characterizing gain spectrum of \acp{EDFA} is one of the key factors to design low margin optical networks, and efficient physical layer control and management. 

The gain spectrum of an \ac{EDFA} has a complex dependence on channel loading, pump power, and operating mode, which makes it difficult to achieve high accuracy with a theoretical model.
Recently, \ac{ML} techniques such as \acp{NN} have been used to build \ac{EDFA} gain models~\cite{zhuMachineLearningBased2018, zhuHybridMachineLearning2020}. Other work~\cite{darosMachineLearningbasedEDFA2020} has produced generalized \ac{ML}-based \ac{EDFA} models using training datasets collected from multiple \acp{EDFA} of the same make and model, which are shown to achieve lower \ac{MAE} of the gain spectrum prediction across multiple devices of the same make. Although these models achieve high prediction accuracy, they do require a large number of measurements, which can be time-consuming and difficult to obtain if the \ac{EDFA} is in a live network. Due to the complexity of the model, deep learning methods such as \ac{NN} also suffer from non-convex training criteria and local minima, which complicate the training process especially with limited number of measurements. 

\ac{TL} is a promising method to reduce the required number of measurements for gain spectrum modeling. \ac{TL} is a machine learning technique to improve the learning in a new task through the transfer of information from a new domain~\cite{zhuangComprehensiveSurveyTransfer2021a}. Specifically, for modeling the gain spectrum of \acp{EDFA}, a base model can be trained on one \ac{EDFA} which can then be retrained to characterize different devices by using a reduced number of measurements from the new device. Recently, it was demonstrated\cite{Wang:23} that a single \ac{EDFA} model can be transferred between different \acp{EDFA} of the same type using only 0.5\% of the entire dataset, showcasing the potential for efficient model transfer in this domain. However, the application of transfer learning across amplifiers of different types (i.e., from an \ac{EDFA} Booster base model towards an \ac{EDFA} Preamp target model) requires further investigation. In addition, work to date has mostly relied on training data from external features, such as input power levels and output gain spectra, which may not fully capture the complex behavior of \acp{EDFA}. 
\footnote{This paper is a preprint of a paper accepted to IEEE Future Networks World Forum~(FNWF) 2024.}
\begin{figure}[t]
    \centering
    \includegraphics[width=\linewidth]{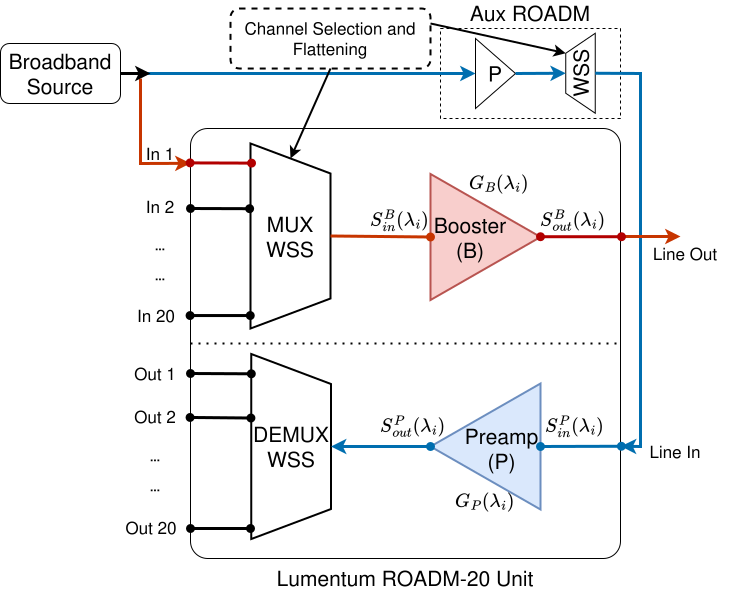}
    \caption{Measurement setup for the Booster/Pre-amplifier EDFA in COSMOS and Open Ireland testbeds.}
    \label{roadm_setup}
\end{figure}

In this paper, we implement and study a novel semi-supervised, self-normalizing \ac{NN} approach (hereafter referred to as the SS-NN model) that characterizes the wavelength-dependent gain of an \ac{EDFA} using just 256 labeled measurements along with additional unlabeled data (which are easier to obtain). By incorporating internal \ac{EDFA} features that are typically available in commercial telecom equipment, our model can be transferred to different \ac{EDFA} types with only a single new measurement through transfer learning. We have reported our SS-NN model previously~\cite{10484559}. In this paper, we have updated the model to perform better in higher error configurations. Furthermore, we describe in more detail the architecture of the SS-NN model, and report the performance analysis. We evaluate our approach on 22 different \acp{EDFA} across the Open Ireland~\cite{open_ireland} (based in Dublin, Ireland) and PAWR COSMOS~\cite{raychaudhuri2020challenge, chenSoftwareDefinedProgrammableTestbed2022} (based in Manhattan, USA) testbeds, achieving a \ac{MAE} within 0.13 dB for same-type transfers and 0.19 dB for cross-type transfers. 

\section{Measurement Setup and Data Collection}
In this section, we describe the experimental setup and data collection strategy of \acp{EDFA} from Open Ireland testbed and PAWR COSMOS testbed. Open Ireland testbed~\cite{open_ireland} is a re-configurable optical-wireless testbed in Dublin, Ireland. PAWR COSMOS testbed~\cite{chenSoftwareDefinedProgrammableTestbed2022, raychaudhuri2020challenge} is a city-scale optical-wireless programmable testbed deployed in Manhattan, USA. 

\subsection{Experimental Setup}

\begin{figure}[t]
    \centering
    \includegraphics[width=0.6\linewidth]{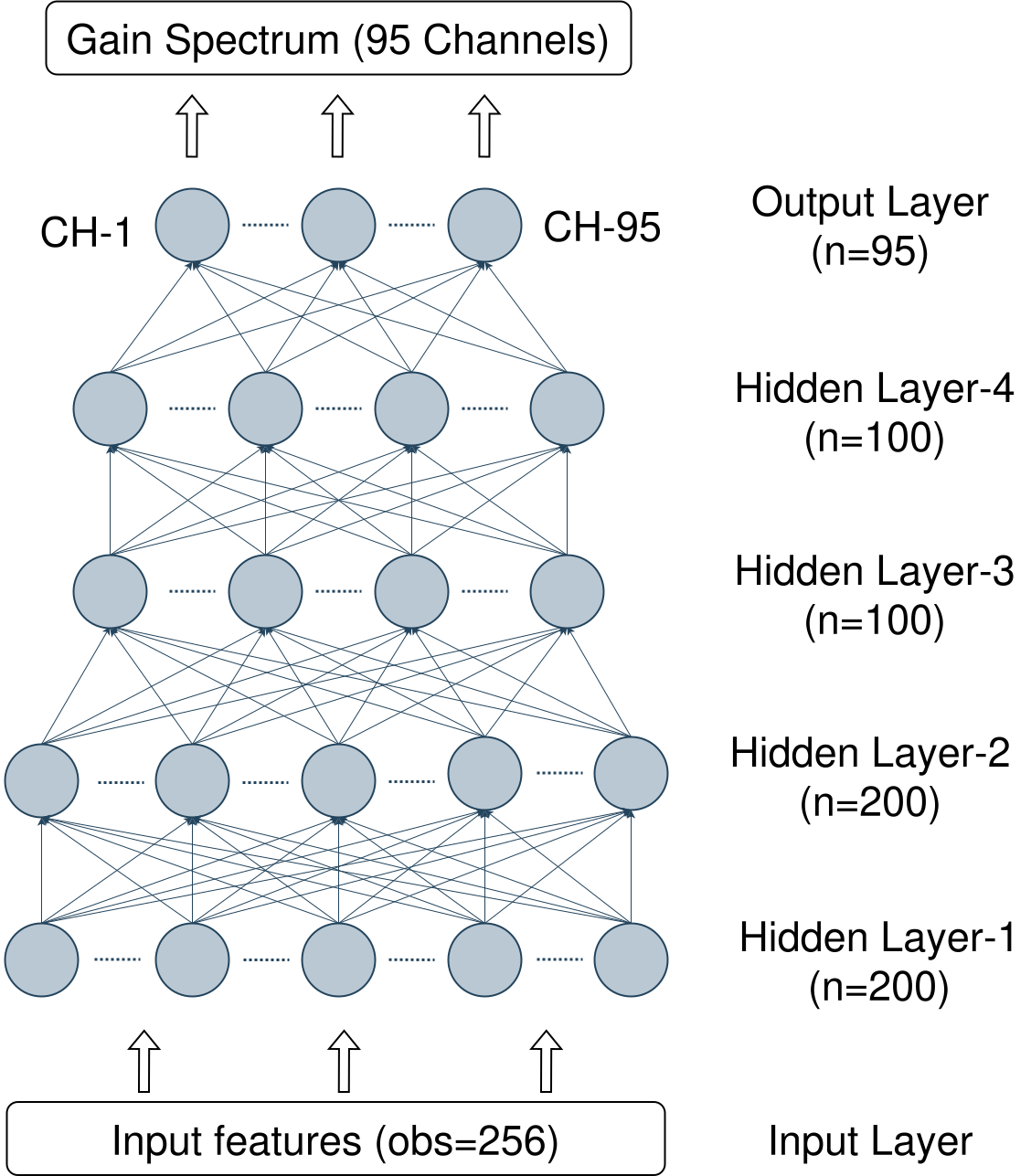}
    \caption{SS-NN model structure with 5 layers.}
    \label{model_structure}
\vspace{-5mm}
\end{figure}

We carry out gain spectrum measurements across multiple wavelengths in the C-band from 3 commercial grade Lumentum ROADM-20 units deployed in the Open Ireland testbed and 8 similar units deployed in the PAWR COSMOS testbed. With each Lumentum ROADM-20 unit containing 2 \acp{EDFA}, we collect data from 11 Boosters and 11 Pre-amplifier \acp{EDFA} in total. 
Figure~\ref{roadm_setup} shows the experimental topology. An \ac{ASE} broadband source is used to generate 95 X 50 GHz \ac{WDM} channels in the C-band according to the \ac{ITU} \ac{DWDM} 50 GHz grid specification. To ensure consistency, we followed a similar measurement setup and data collection pipeline for both testbeds~\cite{wangOpenEDFAGain2023}.

In the data collection for Boosters, the MUX \ac{WSS} is used to flatten the channels, and control the power and channel loading configuration. For preamps, the broadband source output is connected to Line-IN port of an auxiliary \acp{ROADM}, whose DEMUX controls the power and channel loading configuration. The output of this auxiliary ROADM is forwarded to the Line-IN of the \ac{ROADM} under test. The input and output power spectra for each of the 95 channels are collected through the built-in \acp{OCM}. Additionally the total input/output power through the \acp{EDFA} are collected through built-in \acp{PD}. 

\begin{figure*}[t]
    \centering
    \includegraphics[width=\linewidth]{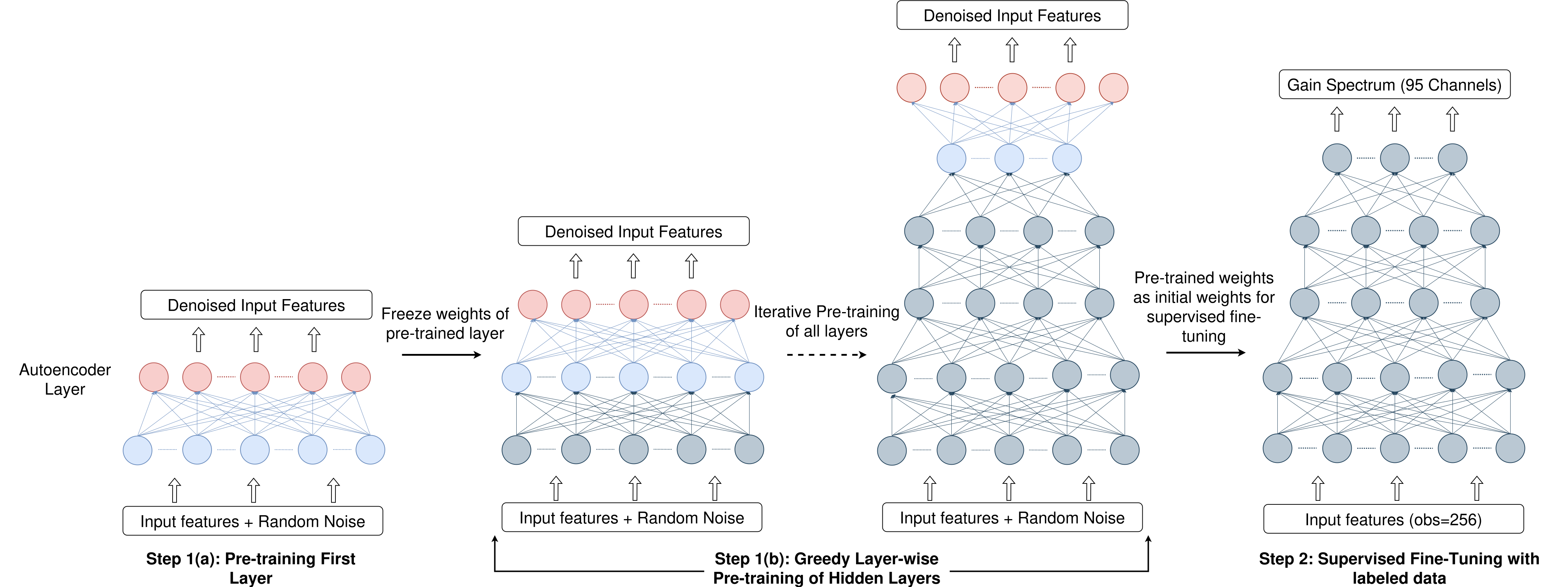}
    \caption{SS-NN model training framework. Step 1(a) and (b) show the greedy layer-wise pretraining of hidden layers using unsupervised pretraining. This pre-trained model forms the basis for Step 2, where supervised fine-tuning is performed with 256 labeled measurements.}
    \label{base_model}
\vspace{-5mm}
\end{figure*}

\subsection{Measurement Configuration}

In the Open Ireland testbed, all \acp{EDFA} were measured at target gains of 15/20/25 dB, while in the COSMOS testbed, the target gains were 15/18/21 dB for Boosters and 15/18/21/24/27 dB for Pre-Amplifiers in high gain mode with 0 dB gain tilt (we adopt different gain setting to emulate diversity of operation in different networks). The dataset includes 3,168 gain measurements (at multiple wavelengths) for each \ac{EDFA}, for each given target gain settings, across 95×50 GHz channels in the C-band, for a total of 202,752 measurements from the COSMOS testbed, and 57,024 measurements from the Open Ireland testbed. In addition, measurements for each \ac{EDFA} are collected under two channel loading modes: Random and Goalpost allocation~\cite{wangOpenEDFAGain2023} (i.e., loading groups of channels in different spectrum bands).

\section{Model Architecture}

In this section, we describe the Semi-Supervised Self-Normalizing Neural Network model for characterizing the gain spectrum of \acp{EDFA}.

\subsection{SNN Model}

%\textcolor{red}{Fig 1 is not described. Also, you should call it a broadband source, as a comb source can be misleading, as this is technically a different type of device}. \textcolor{blue}{Fixed. I have described the topology in Section II (A) Experimental Setup 2nd Paragraph. Please let me know if I should include more details.}
Figure~\ref{model_structure} %\textcolor{red}{(d) should be labeled in the figure} \textcolor{blue}{Fixed. Added a separate figure for highlighting the model structure, and the training process separately.} 
shows the SS-NN model architecture, which consists of an input layer, four hidden layers with 200/200/100/100 neurons, and an output layer. The output layer consists of 95 neurons, predicting the wavelength dependent gain output. The input features to the SS-NN model includes the \ac{EDFA} target gain setting at constant-gain configuration~\((G_0)\), total \ac{EDFA} input power~\((P_{in}\), total \ac{EDFA} output power~\((P_{out}\), input power spectrum~\(\overline{P}{(\lambda_i)} = [P{(\lambda_1)}, P{(\lambda_2)}, P(\lambda_3), ... P{(\lambda_{95})}]\); and a binary vector indicating the channel-loading configuration denoted by~\(\overline{C}=[c_i]_{i=1}^{95}\), with
\vspace{-1mm}
\begin{equation}
    c_i = \begin{cases}
    1, & \text{if the \(i^{th}\) wavelength channel is switched on} \\
    0, & \text{otherwise.}
    \end{cases}
\end{equation}
In addition, we utilize three additional features related to the value of the internal \ac{VOA} in the \ac{EDFA}, namely total \ac{VOA} input and output power~(\(P_{in}^{V}\) and \(P_{out}^{V}\)), and attenuation~\((P_{attn}^{V})\). \acp{VOA} are an internal component of \acp{EDFA}, which indirectly influence the shape of the gain profile by acting on the signal's input powers. This is done to ensure the \ac{EDFA} operates in its design average inversion for a flat spectrum gain profile which matches the \ac{GFF} attenuation~\cite{zyskind_optically_2011}. The \ac{VOA} attenuation is controlled automatically in the \ac{EDFA} based on the model's gain dynamic range, and it grants intrinsic information on the operation of each \ac{EDFA}. Typically, \ac{ML} models for gain spectrum prediction rely only on input and output power spectra information to predict the gain spectrum. However, this choice also treats every \ac{EDFA} like a black box, which leads to poor performance in transfer learning. 

%\textcolor{red}{you should explain why knowledge of the VOA settings is important, and which roles it plays in the EDFA} \textcolor{blue}{Fixed.}

The output layer predicts the gain spectrum~\(\overline{G}{(\lambda_i)} = [G{(\lambda_1)}, G{(\lambda_2)}, G(\lambda_3), ... G{(\lambda_{95})}]\). 

Typically, batch normalization is used to normalize hidden layer outputs~\cite{Wang:23}. However, batch normalization does not perform well when training models with lesser data~\cite{NIPS2017_c54e7837}. Given our objective is to utilize minimal additional measurements for model training, and subsequent transfer learning; we utilize \acp{SNN} with \ac{SELU}~\cite{klambauerSelfNormalizingNeuralNetworks2017a} activation function within layers to render the model as self normalizing. This choice enables us to effectively normalize the hidden layer outputs with a small amount of data, while maintaining the benefits of hidden layer normalization and preserving high accuracy. This step is the key enabling factor of our developed \ac{NN} architecture to achieve effective one-shot training and transferability between models. The \ac{SELU} activation function is given by:
\vspace{-2mm}
\begin{equation}
\text{selu}(x) = \lambda \begin{cases} 
x & \text{if } x > 0 \\ 
\alpha e^x - \alpha & \text{if } x \leq 0 
\end{cases}
\end{equation}

with \(\alpha = 1.673\) and \(\lambda = 1.050\).

\subsection{Training Process}

\begin{figure*}[t]
    \includegraphics[width=\linewidth]{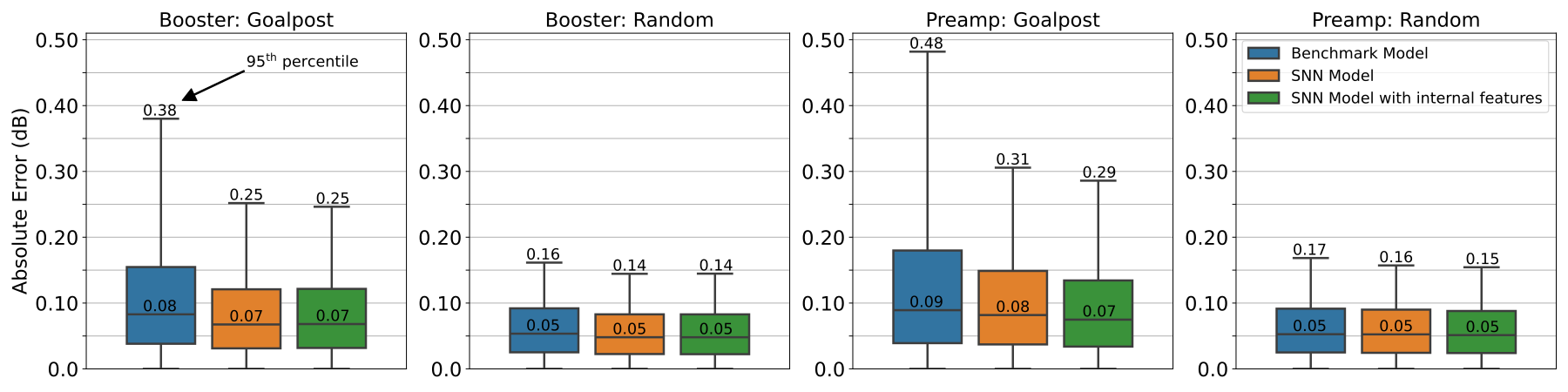}
    \caption{Boxplot distribution of absolute errors across all 11 Booster and 11 Pre-amplifier EDFAs for goalpost and random channel loading. The boxes denote the inter-quartile range, and the whiskers denote the
min/95th percentile}
    \label{boxplot}\
\vspace{-5mm}
\end{figure*}

We use a two-step process to train this model, which includes unsupervised pre-training~\cite{NIPS2006_5da713a6, geProvableAdvantageUnsupervised2023} and supervised fine-tuning~\cite{s22114157}. The training process has been selected with 2 key points: \begin{enumerate}
    \item Unsupervised pre-training utilizes unlabeled data points, which in this case is the input power spectrum of the measurements. These unlabeled measurements are easier to obtain and can also be simulated in cases of flat spectrum cases. This leads to much lower requirement of labeled measurements which are time consuming. 
    \item Unsupervised pre-training leads to a better initial weight initialization than random initialization, and captures more intricate dependency between parameters~\cite{geProvableAdvantageUnsupervised2023}. Additionally, neural networks with pre-training exhibit properties of a regularizer which leads to better generalization~\cite{erhan2010does}. This is especially beneficial for transfer learning from one \ac{EDFA} to other \acp{EDFA}.
\end{enumerate}

Figure~\ref{base_model} shows the training process in detail. In the unsupervised pre-training step, we incrementally initialize the weights of each layer in the model in a greedy manner. First we take 512 unlabeled measurements for each target gain setting. Gaussian noise is added to the measurements. 
We utilize an auto-encoder layer, with the same number of neurons as the dimensions of the feature set, for reconstructing the input features. Starting incrementally from the bottom, the autoencoder layer takes the outputs from the noised inputs, and has the task to predict the denoised inputs. This is achieved by training the layer under test, and autoencoder layer with \ac{MSE} to evaluate how good the model is at reconstructing the input even in the presence of noise. Each layer is trained in a greedy manner for 1,800 epochs with a \ac{LR} of \(1e-03\), along with \ac{MSE} loss function. After the weights of one layer are initialized, its weights are frozen for training of subsequent layer. After all the layers are pre-trained in this manner, the weights are frozen and used as the base model for the next step.

Next is the supervised fine-tuning step, where we utilize 512 measurements to train and fine-tune the model. We use fully and randomly loaded measurements for this step. The model is trained using a modified \ac{MSE} loss function, where the error for any \(k^{th}\) measurement is calculated as below: 
\vspace{-3mm}
\begin{equation}
\label{loss_function}
\text{MSE}_k = \frac{1}{\sum_{i=1}^{95} c_i^k} \cdot \sum_{i=1}^{95} c_i^k . \left[ g_{\text{pred}}^k(\lambda_i) - g_{\text{meas}}^k(\lambda_i) \right]^2
\end{equation}

The model is fine-tuned using Adam Optimizer, with a \ac{LR} of \(1e-03\) over 1,200 epochs, and a gradient clipping threshold of 1.0 for stable training. 

\begin{enumerate}
    \item Usage of less data for training.
    \item Better generalization for transfer learning. 
\end{enumerate}

\vspace{-1mm}
\subsection{Training and Test Sets}

\begin{figure*}[t]

    \centering
    \includegraphics[width=\linewidth]{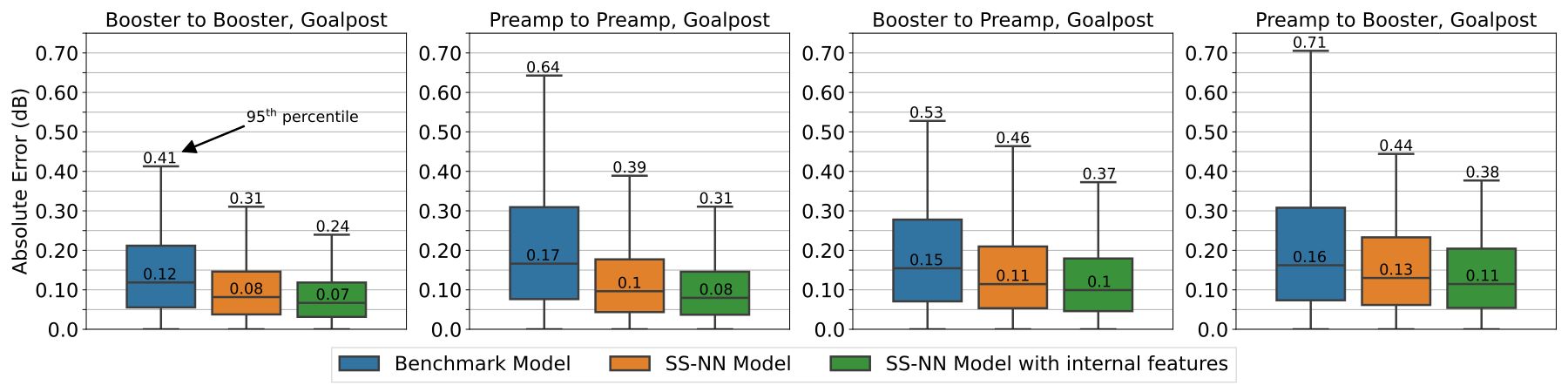}
    \vspace{-9mm}
%     \caption{Boxplot distribution of absolute errors for \textbf{Goalpost} channel loading across all 22 EDFAs for (a) Booster to Booster TL, (b) PreAmp to Preamp TL, (c) Booster to
% Preamp TL and (d) Preamp to Booster TL. The boxes denote the inter-quartile range, and the whiskers denote the min/95th percentile}
%     \label{tl_boxplot_goalpost}

    \centering
    \includegraphics[width=\linewidth]{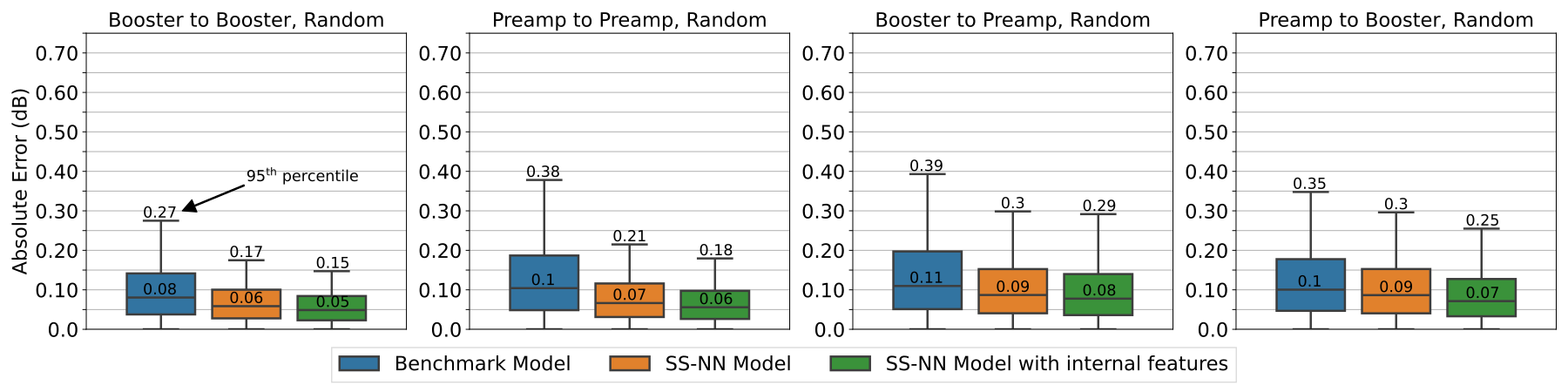}
    \vspace{-8mm}
    \caption{Boxplot distribution of absolute errors across all 22 EDFAs for (a) Booster to Booster TL, (b) PreAmp to Preamp TL, (c) Booster to
Preamp TL and (d) Preamp to Booster TL, for random and goalpost channel loading configurations. The boxes denote the inter-quartile range, and the whiskers denote the min/95th percentile}
    \label{tl_boxplot}
    \vspace{-3mm}
\end{figure*}

We compare the SS-NN model with a benchmark state-of-the-art method~\cite{Wang:23, wangOpenEDFAGain2023}. For equivalent comparison, we follow the same dataset selection criteria. For each gain setting, we split the dataset into a training/test set ratio of 0.86/0.14. The test set contains 436 gain spectrum measurements per gain setting. This test set contains a mixture of random and goalpost channel loading measurements, which represent a diverse set of channel loading configurations. Note that although the SS-NN model uses less data for training, we allocate a larger portion of training data for the benchmark model, which uses 2,732 measurements per gain setting. 

\subsection{SNN Model Performance}

We compare the SS-NN model with the benchmark model using the same set of features to highlight the benefits of our approach. Additionally, we demonstrate the advantage of incorporating internal \ac{EDFA} features by comparing the SS-NN model with and without including these additional features. 

Figure~\ref{boxplot} shows the distribution of absolute errors of gain spectrum predicted by the benchmark model, SS-NN model using same set of features, and SS-NN model with additional internal \ac{VOA} features. The errors are calculated across 11 boosters and pre-amplifier \acp{EDFA} in the Open Ireland and COSMOS testbeds on the test set with random and goalpost channel configurations. For boosters, the SS-NN model achieves a mean absolute error of 0.07 dB and 0.05 dB under the goalpost and random channel configurations. This is comparable to the performance of the benchmark model which uses a considerably higher number of measurements (8196 measurements), compared to a total of 1,792 measurements utilized by the SS-NN model. Importantly, the SS-NN models exhibit a superior error distribution, with a narrow inter-quartile range, and a 95\textsuperscript{th} percentile error of 0.25/0.14 dB, compared to 0.38/0.16 dB by the benchmark model, across the goalpost/random test sets. 

For preamps, the SS-NN model achieves a mean absolute error of 0.08/0.05 dB using the same set of features, and 0.07/0.05 dB using additional internal features across goalpost/random channel configurations. This is marginally better than the benchmark model which achieves a 0.09/0.05 dB error across goalpost/random test sets. Additionally, the distribution of errors for SS-NN models are more stable, with a narrow inter-quartile range, and a 95\textsuperscript{th} percentile error within 0.3 dB across both channel configurations, showing that the SS-NN model generalizes well to unseen channel configurations even when trained with reduced measurements. It should be noted that using additional internal features when directly training \ac{EDFA} models i.e., training on the source \ac{EDFA}'s measurements without \ac{TL} does not provide additional performance. Note that since the resolution of the \ac{OCM} readings is 0.1 dB per channel, the model is achieving the limit of \(\pm\)0.05 dB quantization error in some cases.

% \begin{figure*}[]
%     \centering
%     \includesvg[width=0.49\linewidth,inkscapelatex=false]{figures/edfa_booster_errors.svg}
%     \includesvg[width=0.49\linewidth,inkscapelatex=false]{figures/edfa_preamp_errors.svg}
%     \caption{Distribution of Mean Absolute Error (dB) for Booster and Preamp EDFAs}
%     \label{edfa_errors}

% \end{figure*}

% \begin{table}
%     \centering
%     \renewcommand{\arraystretch}{1.5}
% \caption{}
% \label{tab:my_table}
%     \begin{tabular}{|c|c|c|c|c|}
%     \hline 
%          \textbf{EDFA Type}&  \textbf{ROADM \#}&  \textbf{Mean}&  \textbf{95th Percentile}& \textbf{Maximum}\\ \hline 
%          \multirow{3}{*}{Booster} &  ROADM-1&  &  & \\ \cline{2-5} 
%          &  ROADM-2&  &  & \\  \cline{2-5} 
%          &  ROADM-3&  &  & \\  \hline 
%          \multirow{3}{*}{Preamp}&  ROADM-1&  &  & \\ \cline{2-5} 
%          &  ROADM-2&  &  & \\  \cline{2-5} 
%          &  ROADM-3&  &  & \\ \hline 
%     \end{tabular}
    
% \end{table}

% \begin{figure*}[]
%     \centering
%     \includesvg[width=\linewidth,inkscapelatex=false]{figures/ecdf.svg}
%     \caption{Empirical Cumulative Distribution Function (ECDF) curve}
%     \label{ecdf}

% \end{figure*}

\section{Transfer Learning}

\begin{figure*}[t]

    \centering
    \includegraphics[width=\linewidth]{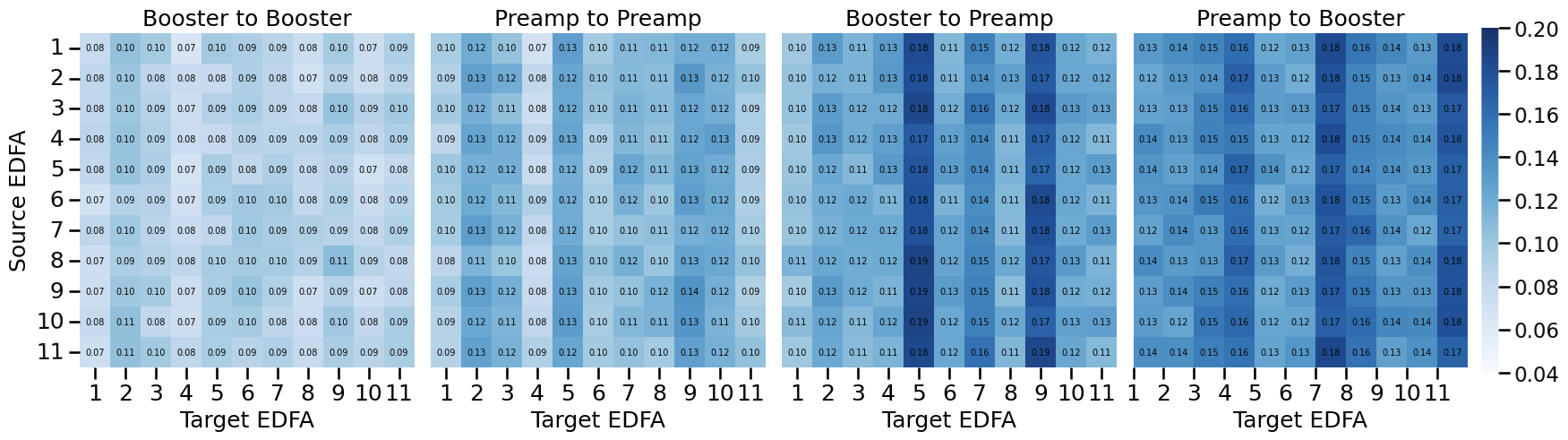}
    \caption{Transfer Learning MAE matrix of SS-NN model with internal features on random loading. The \((i, j)\) entry corresponds to the TL-based EDFA model, where the i\textsuperscript{th} and j\textsuperscript{th} EDFA serve as the source and target models, respectively. EDFA \#1-8 are deployed in COSMOS, while EDFA \#9-11 are deployed in Open Ireland.}
    \label{heatmap}
    \vspace{-5mm}
\end{figure*}

Transfer Learning (TL) is a method to improve the learning in a task through the transfer of information from an existing but related domain. Specifically for \ac{ML} algorithms, \ac{TL} can be used to model a new domain, by transferring a model existing in a related domain~\cite{zhuangComprehensiveSurveyTransfer2021a}. 
\ac{TL} is a viable and useful strategy for modeling the gain spectrum in \acp{EDFA}, reducing the measurement times. In this section, we show that \ac{TL} for SS-NN models can be used to model the gain spectrum across different \acp{EDFA} with minimal additional data. 
\vspace{-1mm}
\subsection{TL Training Process}

To transfer an existing model from a source \ac{EDFA} to a target \ac{EDFA}, we re-train the source model using a single fully-loaded measurement for each target gain setting. This model is trained using the Adam Optimizer for 10,000 epochs using the same \ac{MSE} loss function as Eq.~(\ref{loss_function}) and a gradient clipping threshold of 1.0 for stable training. However, a differential Learning Rate~(LR) is applied across layers instead of a flat \ac{LR}. Specifically, the output layer has a larger \ac{LR} of \(1e-03\) compared to the subsequent hidden layers which have progressively decreasing \acp{LR}, with each layer's rate being 10\% of the next layer's \ac{LR}. In this way, the weights of the output layer are modified more aggressively, allowing it to capture the specific characteristics of the target \ac{EDFA} more effectively. At the same time, the lower levels of the SS-NN model are fine-tuned more gradually to avoid overfitting, ensuring that the model can be generalized to new inputs. 
\vspace{-2mm}
\subsection{Results}

Figure~\ref{tl_boxplot} shows the boxplot of absolute errors of \ac{TL} models for all possible source-target model pairs, for both same type transfer (Booster \(\rightarrow\) Booster, and Preamp \(\rightarrow\) Preamp), as well as cross type transfer (Booster \(\rightarrow\) Preamp, Preamp \(\rightarrow\) Booster); across random and goalpost channel loading configurations. As earlier, we show the comparison for the benchmark model, SS-NN model using same set of features and SS-NN model using additional internal features. 

For the goalpost channel loading configuration, \ac{TL} based SS-NN models achieve a \ac{MAE} less than 0.10/0.13 dB for same-type/cross-type transfers. For the random configuration, the SS-NN models achieve a \ac{MAE} less than 0.07/0.09 dB for same-type/cross-type transfers, respectively. Using additional internal features further improves the performance with a \ac{MAE} less than 0.10/0.13 dB for same-type/cross-type transfers. The \ac{TL} based SS-NN models also outperform the benchmark model, with a better error distribution. 

The results show that SS-NN based \ac{TL} models achieve comparable \ac{MAE} with respect to a directly trained SS-NN model. However, the 95\textsuperscript{th} percentile error of \ac{TL} based models is higher than directly trained models. We believe using more measurements for \ac{TL}, if available, will further improve the performance of \ac{TL} models in high error configurations. It should also be noted that when directly training a SS-NN model on an \ac{EDFA}, including additional internal \ac{VOA} features does not provide much of a performance boost. However, these variables provide a large boost in performance in \ac{TL}, indicating that these extra variables contain distinctive information about behavior of a particular \ac{EDFA}, which improves the performance in \ac{TL}. 

Fig.~\ref{heatmap} shows the \ac{MAE} matrices (in dB) of SS-NN model incorporating internal features across 11 \acp{EDFA} under goalpost channel loading for both same-type and cross-type transfers. In each matrix entry, entry~\((i,i)\) corresponds to a directly trained model (without \ac{TL}), and entry~\((i,j)\) corresponds to the transferred \ac{EDFA} model where the \(i^{th}\) and \(j^{th}\) \ac{EDFA} serve as the source and target models, respectively. The results show that the \ac{TL} performance on each \(j^{th}\) \ac{EDFA} is similar, irrespective of the source \ac{EDFA} model used. The SS-NN based \ac{TL} models achieve a consistent performance for each \ac{EDFA}, close to its directly-trained counterpart even with different source models. Specifically, the SS-NN based \ac{TL} models with internal features under goalpost channel loading achieve a per-\ac{EDFA} \ac{MAE} less than 0.14 dB for same type transfers, and \ac{MAE} less than 0.19 dB for cross type transfers. 

\section{Conclusions}

In this paper, we show a novel Semi-Supervised Self-Normalizing Neural Network (SS-NN) architecture to model the wavelength dependent gain of \acp{EDFA}. The SS-NN model uses a mix of labeled and unlabeled measurements to predict the gain spectrum in a diverse set of channel configurations and target gain settings with high accuracy. Furthermore, the SS-NN model can be transferred to \acp{EDFA} of different types using a single new measurement for each target gain setting with a comparable performance. This demonstrates that a single \ac{EDFA} can be used to characterize multiple \acp{EDFA} using minimal measurements, significantly reducing the amount of data collection. We also find that internal \ac{EDFA} features provide distinctive information about each \ac{EDFA}'s mechanism. Using these internal features provide enchanced performance in both same-type and cross-type \ac{EDFA} transfers, showing potential for improvement by incorporating internal features. We aim to analyze other available internal \ac{EDFA} features, as well as exploring the performance of Transfer Learning when transferring with \acp{ILA} and cross-vendor \acp{EDFA}. 

\bibliographystyle{IEEEtran}
\vspace{-2.5mm}
\bibliography{references.bib}

\vspace{12pt}

\end{document}